\documentstyle[12pt,cite]{article}
\title{QUASI-EXACTLY SOLVABLE POTENTIALS WITH THREE KNOWN EIGENSTATES}
      
\author{T. V. Kuliy$^*$ and V. M. Tkachuk$^{**}$ \\
Ivan Franko Lviv State University, Chair of Theoretical Physics\\ 
		 12 Drahomanov Str., Lviv UA--290005, Ukraine\\
{\small		$^*$E-mail: Kuliy@KTF.Franko.Lviv.UA,
		$^{**}$E-mail: Tkachuk@KTF.Franko.Lviv.UA}}

\begin{document}

\setlength{\jot}{1em}
\setlength{\abovedisplayskip}{1.5em}
\setlength{\belowdisplayskip}{1.5em}

\setcounter{page}{1}
\maketitle
\begin{abstract}
We propose a new SUSY method for the generation of the quasi-exactly
solvable (QES) potentials with three known eigenstates.
New QES potentials and corresponding energy levels 
and wave functions of the ground state and two lowest excited state are 
obtained. A possibility to construct families of exactly solvable 
non--singular potentials which are SUSY partners of the well known ones is 
shown.

\end{abstract}

PACS number(s): 03.65.-w; 11.30.Pb

\section{Introduction}

About twenty years ago an interesting class of the so--called 
quasi--exactly solvable (QES) potentials for which a finite number of 
eigenstates is analytically known was introduced 
\cite{Singh,Fless,Raz,Khare1}. Nowadays the QES problems attract much 
attention 
\cite{Zas1,Zas2,Tub1,Tub2,Shif,Jatkar,Roy1,Gang,Ush,Bagchi,Finkel,Ul,
Vas,Khare2,Bender}. 
Several methods for generation of QES potentials have been worked out and as 
a result many QES potentials have been established. For example three 
different methods based respectively on a polynomial anzatz for wave 
functions, point canonical transformation and supersymmetric (SUSY) quantum 
mechanics are described in the 
paper \cite{Gang}. 

The SUSY is very useful tool for study of exactly solvable potentials. Note 
the papers \cite{Suk,Suk1} where SUSY procedure for constructing 
Hamiltonians either with identical spectra or with identical spectra, apart 
from a missing ground state, was given. This procedure may be repeated again 
and again to generate hierarchies of Hamiltonians whose spectra are related 
to each other. Some recent papers on this subject one can find in 
\cite{Bag1,Bag2,Fer1,Fer2}.
For review of SUSY quantum mechanics see \cite{Lahiri,13}

At the first time the SUSY method for constructing QES potentials was used 
in \cite{Jatkar,Roy1,Gang}. The idea of this method is the following. 
Starting from some initial QES potential with $n+1$ known eigenstates and 
using the properties of the unbroken SUSY one obtains the supersymmetric 
partner potential which is a new QES one with $n$ known eigenstates.

In our previous paper \cite{QES2S} we proposed a new SUSY method for generating 
QES potentials with explicitly known two eigenstates. This method in 
contrast to the the papers \cite{Jatkar,Roy1,Gang} does not require the 
knowledge of the initial QES potential for generating of a new QES one. In 
the present paper we develop this SUSY method for constructing QES 
potentials with explicitly known three eigenstates. 

\section{Witten's model of SUSY quantum mechanics}

Let us first take a look at the Witten's model of SUSY quantum
mechanics. The algebra of SUSY in this case satisfies the following
permutation relations
\begin{eqnarray}\label{1}
&&\{Q^+,Q^-\}=H,\\ \nonumber
&&[Q^\pm,H]=0, \\ \nonumber
&&\left( Q^\pm \right)^2=0, 
\end{eqnarray}
where the supercharges read
\begin{equation} \label{2}
Q^+=B^-\sigma^+, \ \  Q^-=B^+\sigma^-,
\end{equation}
$\sigma^\pm$ are the Pauli matrices,
\begin{equation} \label{3}
B^\pm={1\over\sqrt{2}}\left(\mp{d\over dx}+ W(x)\right),
\end{equation}
$W(x)$ is the superpotential.
The Hamiltonian consists of a pair of standard Schr\"{o}dinger 
operators $H_\pm$
\begin{eqnarray} \label{4}
&& H=\left(
\begin{array} {cc}
{H_+} & {0}\cr
{0} & {H_-}\cr
\end{array}
\right),
\end{eqnarray}
where
\begin{equation} \label{5}
H_\pm=B^\mp B^\pm=-{1\over2}{d^2\over dx^2}+ V_\pm(x),
\end{equation}
$V_\pm(x)$ are the so--called SUSY partner potentials
\begin{equation} \label{6}
V_\pm (x)={1\over 2}\left(W^2(x) \pm W'(x)\right), \ \ W'(x)={dW(x)\over dx}.
\end{equation}
Consider the equation for the energy spectrum
\begin{equation} \label{7}
H_\pm \psi_n^\pm (x)=E_n^\pm \psi_n^\pm (x), \ \
n=0, 1, 2,... .
\end{equation}
The Hamiltonians $H_+$ and $H_-$ have the same energy spectrum
except the zero energy ground state which exists in the case of the unbroken
SUSY. This leads to two--fold degeneracy of the energy spectrum of
$H$ except for the unique zero energy ground state.  Only one of the 
Hamiltonians $H_\pm$ has the zero energy eigenvalue. We shall use the 
convention that the zero energy eigenstate belongs to $H_-$. Due to the
factorization of the Hamiltonians $H_\pm$ (see (\ref{5}))
the ground state for $H_-$ satisfies the equation
\begin{equation} \label{8} 
B^-\psi_0^-(x)=0
\end{equation}
with the solution
\begin{equation} \label{9}
\psi_0^-(x)=C\ \exp\left(-\int W(x) dx\right),
\end{equation}  
here $C$ is the normalization constant.
From the normalization condition it follows that
\begin{equation} \label{10}
{\rm sign}(W(x)) = \pm 1, 
\end{equation}
when
$
x\to \pm\infty. 
$

The eigenvalues and eigenfunctions of the Hamiltonians $H_+$ and $H_-$
are related by the SUSY transformations
\begin{eqnarray} 
&& E_{n+1}^-=E_n^+,\ \ E_0^-=0,  \label{11} \\
&& \psi_{n+1}^-(x)={1\over \sqrt{E_n^+}}B^+\psi_n^+(x), \label{12} \\
&& \psi_n^+(x)={1\over \sqrt{E_{n+1}^-}}B^-\psi_{n+1}^-(x). \label{13}
\end{eqnarray}

The two properties of the unbroken SUSY quantum mechanics, namely, a 
two--fold degeneracy of the spectrum and the existence of the zero energy 
ground state are used for the exact calculation of the energy 
spectrum and wave functions (see reviews \cite{Lahiri,13}).

\section{SUSY constructing QES potentials}\label{QESc}
We shall study the Hamiltonian $H_-$ the ground state of which 
is given by (\ref{9}). Let us consider the SUSY partner of $H_-$, i.e. the 
Hamiltonian $H_+$. If we calculate the ground state of $H_+$ we immediately 
find the first excited state of $H_-$ using the degeneracy of the spectrum 
of the SUSY Hamiltonian and transformations (\ref{11}), (\ref{12}), 
(\ref{13}). In order to calculate the ground state of $H_+$ let us rewrite 
it in the following form
\begin{equation} \label{14}
H_+=H_-^{(1)} + \epsilon, 
\ \  \epsilon > 0,\end {equation}where \begin{eqnarray} \label{15}
&&H_-^{(1)}=B^+_1 B^-_1, \\ \nonumber
&&B^\pm_1={1\over\sqrt{2}}\left(\mp{d\over dx}+ W_1(x)\right),
\end{eqnarray}
$\epsilon$ is the energy of the ground state of $H_+$ 
since $H_-^{(1)}$ has zero energy ground state.

As we see from (\ref{14}), (\ref{15}) the ground state wave function
of $H_+$ is also the ground state wave function of $H_-^{(1)}$ 
and it satisfies the equation
\begin{equation} \label{16}
B^-_1\psi_0^+(x)=0.
\end{equation} 
The solution of this equation is
\begin{equation} \label{17}
\psi_0^+(x)=C\ \exp\left(-\int W_1(x) dx\right).
\end{equation}  

Then  using SUSY transformation (\ref{12}) we can easily calculate the wave 
function of the first excited state of $H_-$. Repeating the described procedure 
for $H_-^{(1)}$ we obtain the second excited state for $H_-$. Continuing 
this procedure $N$ times we obtain $N$ excited states. 
This procedure is well known in SUSY quantum mechanics \cite{Suk,Suk1}
(see also reviews \cite{Lahiri,13}).
The wave functions 
and corresponding energy levels read
\begin{eqnarray}\label{Psin}
\psi_n^-(x)&=&C_n^-B_0^+\ldots B_{n-2}^+ B_{n-1}^+
\exp{\left(-\int W_n(x)dx\right)},\\
\label{En}
E_n^-&=&\sum_{i=0}^{n-1}\epsilon_i,
\end{eqnarray}
where $n=1,2,\ldots,N$. In our notations $\epsilon_0=\epsilon$, 
$B_0^\pm=B^\pm$, $W_0(x)=W(x)$. Operators $B_n^\pm$ are given by (\ref{3})
with the superpotentials $W_n(x)$. The equation (\ref{14}) rewritten for $N$ 
steps 
\begin{equation}\label{Hset}
H^{(n)}_+=H^{(n+1)}_-+\epsilon_n,\ \ \ \ \ 
n=0,1,\ldots,N-1
\end{equation}
leads to the set of equations for superpotentials
\begin{equation}\label{Wset}
W_n^2(x)+W'_n(x)=W^2_{n+1}(x)-W'_{n+1}(x)+2\epsilon_n,\ \ \ \ \ 
n=0,1,\ldots,N-1.
\end{equation}
Previously this set of equations for $W_n(x)$ was solved in the special 
cases of the so--called shape--invariant potentials \cite{SUSY} and 
self--similar potentials for arbitrary $N$ (see review \cite{15}). 
For $N=1$ one can easily obtain a general solution of (\ref{Wset}) without 
restricting ourselves to shape--invariant or self--similar potentials. 
This solution was obtained in \cite{Nik} in the context 
of parasupersymmetric quantum mechanics.  

In our recent paper 
\cite{QES2S} we construct nonsingular solution of (\ref{Wset}) for $N=1$ 
in order to obtain nonsingular QES potentials with two known eigenstates.

In present paper we use the method proposed in \cite{QES2S} to solve the set 
of equations (\ref{Wset}) for $N=2$. It give us the possibility to obtain 
the general expression for wave functions and energy levels of QES 
potentials with three explicitly known eigenstates. 

Let us introduce new functions
\begin{eqnarray}
&&W_+^{(n)}(x)=W_{n+1}(x)+W_n(x),\\\nonumber
&&W_-^{(n)}(x)=W_{n+1}(x)-W_n(x),\ \ \ n=0,1,\ldots,N-1.
\end{eqnarray}
Then equations (\ref{Wset}) read
\begin{eqnarray}
{W_+'}^{(n)}(x)=W_-^{(n)}(x)W_+^{(n)}(x)+2\epsilon_n.
\end{eqnarray}
One can easily solve these equations with respect to the functions 
$W_-^{(n)}(x)$ obtaining the following expressions for superpotentials
\begin{eqnarray}  \label{Wneq}
&&W_n(x)={1\over 2}\left(W_+^{(n)}(x) - 
{{W'^{(n)}_+(x)-2\epsilon_n}\over W_+^{(n)}(x)} \right), 
\\ \nonumber
&&W_{n+1}(x)={1\over 2}\left(W_+^{(n)}(x) + 
{{W'^{(n)}_+(x)-2\epsilon_n}\over W_+^{(n)}(x)} \right), 
\end{eqnarray}
which lead obviously to the following set of equations for functions 
$W_+^{(n)}(x)$
\begin{eqnarray}  \label{Wpset}
&&W^{(n)}_+(x) + {{W'^{(n)}_+(x)-2\epsilon_n}\over W^{(n)}_+(x)}
=
W^{(n+1)}_+(x) + {{W'^{(n+1)}_+(x)-2\epsilon_{n+1}}\over W^{(n+1)}_+(x)},
\\\nonumber
&&n=0,1,\ldots,N-2.
\end{eqnarray}

Thus the set of $N$ equations (\ref{Wset}) is reduced to the set of $N-1$ 
equations (\ref{Wpset}). In the simplest case $N=1$ equation (\ref{Wpset}) 
is absent and relations (\ref{Wneq}) express just a general solution of 
equation (\ref{Wset}). 
To obtain a general solution of set (\ref{Wset})
in the case $N=2$ we have to solve one equation:
\begin{eqnarray}  \label{Wp-Wp1}
W_+(x) + {{W'_+(x)-2\epsilon}\over W_+(x)}=
\tilde{W}_+(x) - {{{\tilde{W}_+}'(x)-2\epsilon_1}\over \tilde{W}_+(x)},
\end{eqnarray}
where we have introduced for short the notation
\begin{eqnarray}  
W_+(x)\equiv W^{(0)}_+(x),&&\tilde{W}_+(x)\equiv W^{(1)}_+(x).
\end{eqnarray}
It is easily to rewrite this equation as follows
\begin{eqnarray}  \label{Wp-Wp1-r}
W_+(x)\tilde{W}_+(x)(\tilde{W}_+(x)-W_+(x))-(W_+(x)\tilde{W}_+(x))' 
+ 2(\epsilon_1 W_+(x)+\epsilon \tilde{W}_+(x))=0,
\end{eqnarray}
or
\begin{eqnarray}  \label{Wp-U}
U(x)\left({U(x)\over W_+(x)}-W_+(x)\right)-U'(x) 
+ 2\left(\epsilon_1 W_+(x)+\epsilon {U(x)\over W_+(x)}\right)=0,
\end{eqnarray}
where we have introduced the function 
\begin{equation}
U(x)=W_+(x)\tilde{W}_+(x).
\end{equation}
We again came to the Riccati equation with respect to $U(x)$.
On the other hand it is an algebraic equation with respect to 
$W_+(x)$. Thus we can start from arbitrary function $U(x)$ to 
construct the functions $W_+(x)$ and $\tilde{W}_+(x)$ which take the form
\begin{eqnarray}  \label{Wp}
W_+(x)={2 U(x)(U(x)+2\epsilon)\over 
U'(x)\left(1+
{\cal R}(x)
\right)},\\\nonumber
\tilde{W}_+(x)={U'(x)\left(1+
{\cal R}(x)
\right)\over 2 (U(x)+2\epsilon)},
\end{eqnarray}
where
\begin{eqnarray}\label{cR}
{\cal R}(x)&=&\pm R(x),\\\label{R}
R(x)&=&\sqrt{1+4{U(x)(U(x)+2\epsilon)(U(x)-2\epsilon_1)\over 
U'(x)^2}}.
\end{eqnarray}
We mean square root $R(x)$ as positively defined value, while function 
${\cal R}(x)$ can be chosen in the form of $R(x)$ or $-R(x)$ within 
different intervals separated by zeros of function $R(x)$. Note that just 
the possibility to choose different signs allows us, as will be shown in the 
section \ref{exact}, to construct in a simple way new exactly solvable 
potentials using the known ones. 

Now we can obtain 
three consequent superpotentials $W(x)$, $W_1(x)$ and $W_2(x)$ using the 
relations (\ref{Wneq}).
Then using (\ref{Psin}) and (\ref{En}) we obtain the energy levels
and the wave functions of the first and the second excited states
for $H_-$ 
\begin{eqnarray}
&&E_1^-=\epsilon,\ \ \ \
E_2^-=\epsilon+\epsilon_1,\label{E12}\\
\label{Psi1}
&&\psi_1^-(x)=C_1 W_+(x)
\exp\left(-\int W_1(x) dx\right),\\\nonumber&&
\psi_2^-(x)=C_2\left((W(x)+W_2(x))\tilde{W}_+(x)-\tilde{W}'_+\right)
\exp\left(-\int W_2(x) dx\right).
\end{eqnarray}

The superpotentials $W_1(x)$ and $W_2(x)$ must satisfy the same condition  
(\ref{10}) as $W(x)$. It leads to the same limitations for the functions 
$W_+(x)$ and $\tilde{W}_+(x)$. Both must be positive at infinity, negative 
at minus infinity and therefore each of them must posses at least one zero.
Let us consider at first continuous superpotentials. As is  seen from the 
(\ref{Wneq}) to avoid singularity of superpotentials the 
functions $W_+(x)$ and $\tilde{W}_+(x)$ ought have each only one zero 
\cite{QES2S} 
$$ 
W_+(x_0)=0,\ \ \ \ \tilde{W}_+(\tilde{x}_0)=0
$$
at which they must satisfy the condition
\begin{eqnarray}\label{Wp-cond}
W'_+(x_0)=2\epsilon,\ \ \ \ {\tilde{W}_+}'(\tilde{x}_0)=2\epsilon_1.
\end{eqnarray}

Thus we have a number of limitations in the choice of function $U(x)$ as a 
product of $W_+(x)$ and $\tilde{W}_+(x)$.
There are two different possibilities of the choice of the function $U(x)$.
Either $x_0=\tilde{x}_0$ that provides $U(x)$ to have only one second--order 
zero point and to be positive at all the rest number line 
\begin{eqnarray}\label{U1}
\left\{
\begin{array}{lll}
U(x_0)=0,&U'(x_0)=0,&U''(x_0)>0, \cr
U(x)>0,& x\ne x_0, \cr
\end{array} 
\right.
\end{eqnarray}
or $x_0\ne \tilde{x}_0$ and therefore $U(x)$ has two zero points and changes 
its sign as follows 
\begin{eqnarray}\label{U2}
\left\{
\begin{array}{ll}
U(x)<0,& x \in ({\rm min}[x_0,\tilde{x}_0],{\rm max}[x_0,\tilde{x}_0]),\cr
U(x_0)=U(\tilde{x}_0)=0,& U'({\rm min}[x_0,\tilde{x}_0])<0,\ \ \ 
U'({\rm max}[x_0,\tilde{x}_0])>0,\cr
U(x)>0,& x \notin [{\rm min}[x_0,\tilde{x}_0],{\rm max}[x_0,\tilde{x}_0]].\cr
\end{array} 
\right.
\end{eqnarray}

The sign of the function ${\cal R}(x)$ in the expressions (\ref{Wp}) for the 
functions 
$W_+(x)$ and $\tilde{W}_+(x)$ should be chosen in a such way to ensure 
smoothness of these functions and an existence of one zero for each of them. 
A full analysis of the conditions which must satisfy the function 
$U(x)$ to provide continuous superpotentials is rather boring and 
includes consideration of the behaviour of the functions 
$W_+(x)$ and $\tilde{W}_+(x)$ nearby possible 
zeroes of the expressions
\begin{equation}
U(x)+2\epsilon,\ \ \ U(x)-2\epsilon_1,\ \ \ U'(x), \ \ \ R(x),
\end{equation}
which is crucial for the continuity of the final superpotentials
$W(x)$, $W_1(x)$ and $W_2(x)$.

We shall consider more closely the simplest 
case of functions $U(x)$ which has one zero point and satisfies the 
conditions (\ref{U1}). The other condition which the function $U(x)$ must 
satisfy is a consequence of the conditions (\ref{Wp-cond}) that connect the 
derivatives of the functions $W_+(x)$ and $\tilde{W}_+(x)$ with the energies 
$\epsilon$ and $\epsilon_1$. One can easily find 
that
\begin{eqnarray}\label{U''}U''(x_0)&=&W_+''(x_0)\tilde{W}_+(x_0)+2 
W_+'(x_0){\tilde{W}_+}'(x_0)+W_+(x_0){\tilde{W}_+}''(x_0)\\\nonumber&=&2 
W_+'(x_0){\tilde{W}_+}'(x_0)=8\epsilon\epsilon_1.
\end{eqnarray}
The other obvious condition imposed on the function $U(x)$
is positivity of the expression under the square root of the 
function $R(x)$ (\ref{R})
\begin{equation}\label{Sq}
{U'(x)^2+4 U(x)(U(x)+2\epsilon)(U(x)-2\epsilon_1)
\over 
U'(x)^2}
\ge0.
\end{equation}
One can easily check that $R(x_0)=0$. Let us consider the case 
when point $x_0$ is unique zero of the function $R(x)$. Then the only way to 
construct non--singular potentials is to chose function 
${\cal R}(x)=R(x)$ over all the line in (\ref{cR}). Besides we shall require 
the function $R(x)$ to be smooth in the vicinity of point $x_0$ to avoid 
cusps of functions $W_+(x)$ and $\tilde{W}_+(x)$ at this point. Note that 
such cusps would result in $\delta$--like singularities of the final 
potential $V_-(x)$.

Thus we obtain one of possible sets of conditions for the 
function $U(x)$ allowing to construct non--singular QES potentials with 
three known eigenstates 
\begin{eqnarray}\label{U-cond}
&&U(x)>0 \ \ \forall \ x\ne x_0,\\\nonumber
&&U(x_0)=0, \ \ \ 
U'(x_0)=0, \ \ \
U''(x_0)=8\epsilon\epsilon_1, \ \ \ 
U'''(x_0)=0, \\\nonumber
&&
U^{(4)}(x_0)=64\epsilon\epsilon_1(\epsilon_1-\epsilon), \ \ \
U^{(5)}(x_0)=0, \ \ \ 
{U^{(6)}(x_0)\over 8\epsilon\epsilon_1}\ge
32(2\epsilon_1^2-13\epsilon\epsilon_1+2\epsilon^2),\\\nonumber
&&R(x)>0 \ \ \forall \ x\ne x_0.
\end{eqnarray}

Note that the most of exactly solvable potentials which are continuous 
over all the line satisfy these conditions. 

\section{Examples}
One can easily check that the simplest functions $U(x)$ yield us 
the well--known potentials. 
For example starting from 
$U(x)=4\epsilon\epsilon_1x^2$ at  $\epsilon_1=\epsilon$
we get harmonic oscillator potential.
Another simple function 
$U(x)=4\epsilon\epsilon_1\tanh^2 x$ at $\epsilon_1=\epsilon-1$
leads us to the well--known exactly solvable Rosen--Morse potential.
One more simple example
$U(x)=4\epsilon\epsilon_1\sinh^2 x$ at $\epsilon_1=\epsilon+{1\over 2}$
reproduces the special case of the well--known quasi exactly solvable 
Razavy potential \cite{Raz}.

Let us consider more complicated examples leading to new QES potentials. We 
shall start from the function\begin{eqnarray}\label{U-ex0}
U(x)=4\epsilon\epsilon_1x^2{1+a^2x^2\over 1+b^2x^2},&& 
\end{eqnarray}
where $a$ and $b$ are real parameters.

Due to the conditions for $U^{(4)}(x_0)$ and $U^{(6)}(x_0)$
from (\ref{U-cond}) we get
\begin{eqnarray}
a^2=b^2+{2\over 3}(\epsilon_1-\epsilon)\label{EpEpab}
\end{eqnarray}
and
\begin{eqnarray}\label{EpEp}
{13\over 4}{\epsilon\over b^2}
	-{15\over 8}-{3\over 8}\sqrt{\Delta}
\le{\epsilon_1\over b^2}\le
{13\over 4}{\epsilon\over b^2}
	-{15\over 8}+{3\over 8}\sqrt{\Delta}
\end{eqnarray}
respectively; here
\begin{equation}
	\Delta=
	{25-60{\epsilon\over b^2}+68\left({\epsilon\over b^2}\right)^2}.
\end{equation}

Because $a$ and $b$ are real numbers it follows from 
(\ref{EpEpab}) that
\begin{equation}
{\epsilon\over b^2}-{3\over 2} \le {\epsilon_1\over b^2}.
\end{equation}
It is easily to check that
\begin{equation}
{13\over 4}{\epsilon\over b^2}
	-{15\over 8}-{3\over 8}\sqrt{\Delta}
\le
{\epsilon\over b^2}-{3\over 2}
\end{equation}
and therefore (\ref{EpEp}) together with (\ref{EpEpab}) 
lead to the following inequality
\begin{eqnarray}\label{Epfin}
{\epsilon\over b^2}-{3\over 2}\le{\epsilon_1\over b^2}\le
{13\over 4}{\epsilon\over b^2}
	-{15\over 8}+{3\over 8}\sqrt{\Delta}.
\end{eqnarray}

Thus in order to obtain nonsingular potentials the parameters of 
function (\ref{U-ex0}) must satisfy conditions (\ref{EpEpab}) and 
(\ref{Epfin}). Besides we obviously must require the parameters  $\epsilon$ 
and $\epsilon_1$ to be positive. 

We shall omit the general expression for the potential $V_-(x)$ as it is 
huge and rather useless. It is easily to show that there are only two sets 
of positive $\epsilon$ and $\epsilon_1$ allowing to resolve the root in 
the function $R(x)$ (Cases 1 and 2) and therefore to simplify significantly 
the final expressions. The other simplified expression (Case 3) we shall 
obtain putting $a=0$ that corresponds to the lowest value of the parameter 
$\epsilon_1$ in inequality (\ref{Epfin}).

\subsection{Case 1}
In the case   
\begin{eqnarray}
\epsilon={3\over 2}b^2,&&\epsilon_1=\left({3\over 2}+\sqrt{3}\right)b^2
\end{eqnarray}
square root in $R(x)$ can be resolved and we obtain the potential
\begin{eqnarray}
V_-(x)&=&
\left(-{9\over 4}-2 {\sqrt{3}}+ 
3\left({7\over 8}+{\sqrt{3}\over 2}\right)b^2{x^2}+\right.\\\nonumber
&&\left.
{3-\sqrt{3}\over 2-{\sqrt{3}}+b^2 x^2}+
{{7\sqrt{3}-12}\over{{\left(2-{\sqrt{3}}+b^2 x^2\right)}^2}}\right)b^2
\end{eqnarray}
and the eigenfunctions
\begin{eqnarray}
\psi_0(x)&=&
C{e^{-{\left(3+2 {\sqrt{3}}\right)\over 4}b^2 x^2}} 
{{\left(2-\sqrt{3}+b^2 
x^2\right)}^{{\left(3-{\sqrt{3}}\right)\over 2}}},\\ 
 \psi_1(x)&=&
C_1{e^{-{\left(3+2\sqrt{3}\right)\over 4}b^2 x^2}}  
{{\left(2-\sqrt{3}+b^2 x^2\right)}^{{\left(\sqrt{3}-1\right)\over 2} 
}} x,\\
\psi_2(x)&=&
C_2{e^{-{\left(3+2\sqrt{3}\right)\over 4}b^2 x^2}} 
{{\left(2-\sqrt{3}+b^2 x^2\right)}^{\left(\sqrt{3}-1\right)\over 2}}
\left(2-\sqrt{3}-b^2 x^2\right)
\end{eqnarray}
corresponding to the three lowest levels
\begin{eqnarray}
E^-_0=0,\ \ \ \ E^-_1={3\over 2}b^2,\ \ \ \ E^-_2=\left(3+\sqrt{3}\right)b^2. 
\end{eqnarray}

To simplify these expressions let us make the substitution
\begin{eqnarray}
b^2=\left(2-\sqrt{3}\right)c^2.
\end{eqnarray}
Thus we get
\begin{eqnarray}\label{V1}
V_-(x)&=&
c^2\left( 
{3\over 8}c^2 x^2+
{3-\sqrt{3}\over 1+c^2 x^2}+
{{2\sqrt{3}-3}\over{{\left(1+c^2 x^2\right)}^2}}
-{7\sqrt{3}\over 4}+{3\over 2}\right),\\
E^-_0=0,&& E^-_1=3\left(1-{\sqrt{3}\over 2}\right)c^2,\ \ \ \ 
E^-_2=\left(3-\sqrt{3}\right)c^2,\\
\psi^-_0(x)&=&
C{e^{-{\sqrt{3}\over 4}c^2 x^2}} 
{{\left(1+c^2 x^2\right)}^{{\left(3-{\sqrt{3}}\right)\over 2}}},\\ 
 \psi^-_1(x)&=&
C_1{e^{-{\sqrt{3}\over 4}c^2 x^2}}  
{{\left(1+c^2 x^2\right)}^{{\left(\sqrt{3}-1\right)\over 2} 
}} x,\\
\psi^-_2(x)&=&
C_2{e^{-{\sqrt{3}\over 4}c^2 x^2}} 
{{\left(1+c^2 x^2\right)}^{\left(\sqrt{3}-1\right)\over 2}}
\left(1-c^2 x^2\right).
\end{eqnarray}

It is worth to note that the obtained potential 
(\ref{V1}) is double--well one.
As well known, double--well potentials has been used extensively 
to model a wide range of natural phenomena. 

\subsection{Case 2}
The other set of positive $\epsilon$ and $\epsilon_1$  
resolving the root in the function $R(x)$ reads 
\begin{eqnarray}
\epsilon={3\over 2}b^2,&&\epsilon_1={1\over 2}b^2.
\end{eqnarray}
It leads to the well--known supersymmetric partner of a harmonic oscillator
\cite{Bag1,Bag2,Jun}
\begin{eqnarray}\label{case2}
V_-(x)&=&
{3b^2\over 4}+{b^4 x^2\over 8}-{4b^2\over{(1+b^2 x^2)^2}}
+{2b^2\over 1+b^2 x^2}.
\end{eqnarray}
This potential is exactly solvable although we have got it using the 
procedure for constructing QES potentials with three exactly known 
eigenstates. We will also obtain the same potential in the section 
\ref{exact} using the other procedure allowing to construct
new exactly solvable potentials (see (\ref{Vp1osc})).

\subsection{Case 3}
Let us consider another particular case of the potential under consideration 
which is rather simple. We put $a=0$ to reduce the function $U(x)$ 
(\ref{U-ex0}) to the form 
\begin{eqnarray}\label{U-ex2}
U(x)=4\epsilon\epsilon_1{x^2\over 1+b^2x^2}.
\end{eqnarray}
Due to the
relation (\ref{EpEpab}) we obtain the connection between  the energies 
$\epsilon$ and $\epsilon_1$ 
\begin{eqnarray}\label{E1E2-ex2}
\epsilon=\epsilon_1+{3\over 2}b^2 .
\end{eqnarray}
Note that in accordance with relation (\ref{Epfin}) such a choice of 
parameters corresponds to the minimal value of $\epsilon_1$ at given 
$\epsilon$ and $b$.
It leads directly to the potential
\begin{eqnarray}\label{V3}
{V_-(x)\over b^2}&=&
{
(\alpha^2+3\alpha+1)^2
\over 
8\alpha(\alpha+1)
}+
{1\over 1+{b^2} {x^2}}-
{2\over(1+b^2 x^2)^2}-{2\over\rho(x){{(1+{b^2} {x^2})}^2}}\\\nonumber
&&-
{\alpha^2+\alpha-1
\over
\rho(x)(1+b^2 x^2)
}
-
{
(\alpha^2+\alpha+1)^3
\over 
8\alpha(\alpha+1)\rho^2(x)
}
+{(\alpha^2+\alpha+1)^2\over 4\rho^3(x)}
\end{eqnarray}
and the following eigenfunctions
\begin{eqnarray}\label{3Psi0}
\psi^-_0(x)&=&C_0
{
\rho(x)+\alpha
\over
\rho(x)-1
}
e^{
-{1\over 2}\rho(x)
\left(
	1+
	{1
	\over
	\alpha
	}
	+
	{1
	\over 
	\alpha+1
	}
\right)},
\\\label{3Psi1}
\psi^-_1(x)&=&C_1
{
x
\over
\rho(x)-1
}
e^{
-{1\over 2}\rho(x)
\left(
	1-
	{1
	\over
	\alpha
	}
	+
	{1
	\over 
	\alpha+1
	}
\right)},
\\\label{3Psi2}
\psi^-_2(x)&=&C_2
{
\rho(x)-\alpha-1
\over
\rho(x)-1
}
e^{
-{1\over 2}\rho(x)
\left(
	1-
	{1
	\over
	\alpha
	}
	-
	{1
	\over 
	\alpha+1
	}
\right)},
\end{eqnarray}
corresponding to the levels
\begin{eqnarray}
E^-_0=0,\ \ \ \ E^-_1=\epsilon_1+{3\over 2}b^2,\ \ \ \ E^-_2=2\epsilon_1+{3\over 2}b^2,
\end{eqnarray}
where
\begin{eqnarray}
\nonumber
&&\rho(x)=\sqrt{1+\alpha(\alpha+1)(1+b^2 x^2)},\ \ \
\alpha=1+2{\epsilon_1\over b^2}.
\end{eqnarray}

Obtained potential (\ref{V3}) has one minimum at $x=0$ and tends to constant 
for $x\to\pm\infty$.
For the case $\alpha>1$ all the obtained wave functions 
(\ref{3Psi0}--\ref{3Psi2}) are square integrable and there exist at least 
three bound state in the well. For $(\sqrt{5}-1)/2<\alpha\le 1$ only two 
lower wave functions (\ref{3Psi0}),(\ref{3Psi1}) are square integrable and 
we have just two bound states, while for $\alpha\le (\sqrt{5}-1)/2$ only the 
ground state wave function (\ref{3Psi0}) remains square integrable and the 
potential has only one bound state.

\section{Constructing exactly solvable potentials}
\label{exact}

Although we have developed our scheme to construct new QES potentials it 
seems to be also of use for constructing SUSY partner 
potentials for the exactly solvable ones. Let us start from some exactly 
solvable potential $V_-(x)$ \begin{equation}
V_- (x)={1\over 2}\left(W^2(x)-W'(x)\right),
\end{equation}
for which we know three first superpotentials $W(x)$, $W_1(x)$ and 
$W_2(x)$ satisfying set of equations (\ref{Wset}). 
One can easily construct the functions $W_+(x)=W(x)+W_1(x)$, 
$\tilde{W}_+(x)=W_1(x)+W_2(x)$ and $U(x)=W_+(x)\tilde{W}_+(x)$. Substituting 
function $U(x)$ into the expressions (\ref{Wp}) we can obviously reproduce 
the functions $W_+(x)$ and $\tilde{W}_+(x)$ choosing corresponding 
signs in the function (\ref{cR}) which we shall denote as ${\cal R}_0(x)$.
Beside the functions $W_+(x)$ 
and $\tilde{W}_+(x)$ we can obviously obtain another pair of functions 
${\cal W}_+(x)$ and $\tilde{\cal W}_+(x)$ given by the same expressions 
(\ref{Wp}) with the only difference that here we choose the function 
(\ref{cR}) with the opposite sign to ${\cal R}_0(x)$, 
\begin{eqnarray}
{\cal R}(x)=-{\cal R}_0(x)
\end{eqnarray}
The new functions ${\cal W}_+(x)$ and $\tilde{\cal W}_+(x)$ satisfy the 
same equation (\ref{Wp-Wp1}) and just they allow us to 
construct new exactly solvable potentials.

Let us consider these functions in more detail.
It is easily to show that both functions 
${\cal W}_+(x)$ and $\tilde{\cal W}_+(x)$
are negative at the infinity and positive at the minus infinity.
Explicit calculations show that it provides 
the same behaviour of the superpotentials ${\cal W}(x)$, ${\cal W}_1(x)$ and 
${\cal W}_2(x)$ which can be obtained by substitution functions ${\cal 
W}_+(x)$ and $\tilde{\cal W}_+(x)$ into relations (\ref{Wneq}). We shall omit 
explicit expressions for the superpotentials ${\cal W}_1(x)$ and ${\cal 
W}_2(x)$ noting only that both of them are singular while the superpotential
\begin{equation} \label{calW}
{\cal W}(x)=
{U'(x)({\cal R}_0(x)-1)\over 2(2\epsilon+U(x))}+
{
8\epsilon\epsilon_1-U''(x)+2U(x)(4(\epsilon_1-\epsilon)-3U(x))
\over 
2U'(x){\cal R}_0(x)
}
\end{equation}
has no singularities if only the initial potential $V_-(x)$ is 
non--singular. 
Using superpotential ${\cal W}(x)$ we can construct in a standard way the 
pair of the Hamiltonians ${\cal H}_-$ and ${\cal H}_+$. Their properties 
will be very similar to that of the Hamiltonians $H_-$ and $H_+$ 
with the only difference that now Hamiltonian ${\cal H}_+$ will have 
zero--energy ground state with corresponding eigenfunction
\begin{equation} 
\label{Phi0}
\varphi^+_0(x)=C\exp\left(\int {\cal W}(x) dx\right).
\end{equation}
All the higher eigenvalues of Hamiltonians ${\cal H}_+$ and ${\cal H}_-$ 
will coincide and corresponding eigenfunctions will be connected as follows
\begin{eqnarray}\label{PhiPhi} 
\varphi_{n+1}^+(x)&=&{1\over \sqrt{2{\cal E}_n^-}}
\left({d\over dx}+{\cal W}(x)\right)\varphi_n^-(x), 
\\\nonumber
\varphi_n^-(x)&=&{1\over \sqrt{2{\cal E}_{n+1}^+}}
\left(-{d\over dx}+{\cal W}(x)\right)\varphi_{n+1}^+(x).
\end{eqnarray}

Let us consider few the simplest explicit examples. In the case of a 
harmonic oscillator for which all the superpotentials and distances between
the energy levels read
\begin{eqnarray}
W_n(x)=\epsilon x,&& \epsilon_n=\epsilon
\end{eqnarray}
we get corresponding function
\begin{eqnarray}
U(x)=4\epsilon^2 x^2
\end{eqnarray}
and we obtain 
\begin{eqnarray}
{\cal R}_0(x)=R(x)=2\epsilon x^2.
\end{eqnarray}
Choosing now ${\cal R}(x)=-{\cal R}_0(x)$ and using (\ref{calW}) we get
\begin{eqnarray}
{\cal W}(x)=-\epsilon x{5+2\epsilon x^2\over 1+2\epsilon x^2},
\end{eqnarray}
leading to the following SUSY partner 
potentials
\begin{eqnarray}\label{Vm1osc}
{\cal V}_-(x)&=&{\epsilon^2 x^2\over 2}+{5\epsilon\over 2},\\\label{Vp1osc}
{\cal V}_+(x)&=&{\epsilon^2 x^2\over 2}
+{4\epsilon\over 1+2\epsilon x^2}
-{8\epsilon\over (1+2\epsilon x^2)^2}
+{3\epsilon\over 2}.
\end{eqnarray}
Thus we have obtained the exactly solvable potential (\ref{Vp1osc}) 
which is a SUSY partner to the harmonic oscillator potential (\ref{Vm1osc}). 
This potential (\ref{Vp1osc}) after substituting $\epsilon=b^2/2$ coincides 
with potential (\ref{case2}) obtained in the section \ref{QESc}.
Let us remind that the potential ${\cal V}_+(x)$ has 
zero--energy level with corresponding wave function (\ref{Phi0}). Therefore 
we can now treat the potentials ${\cal V}_-(x)$ and ${\cal V}_+(x)$ as 
$V_+(x)$ and $V_-(x)$ respectively, that is $V_+(x)={\cal V}_-(x)$ and 
$V_-(x)={\cal V}_+(x)$. The superpotential corresponding to the pair 
$V_-(x)$ and $V_+(x)$ is $W(x)=-{\cal W}(x)$. Because the upper SUSY partner 
$V_+(x)$ is a harmonic oscillator we can easily build up all the hierarchy of 
superpotentials satisfying equations (\ref{Wset}):
\begin{eqnarray}\label{Wn1osc}
W_0(x)=\epsilon x{5+2\epsilon x^2\over 1+2\epsilon x^2},&&W_n(x)=\epsilon x, 
\ \ \ n=1,2,\dots
\end{eqnarray}
The corresponding distances between the energy levels read
\begin{eqnarray}
\epsilon_0=3\epsilon,&&\epsilon_n=\epsilon, \ \ \ n=1,2,\dots
\end{eqnarray}

The obtained potential (\ref{Vp1osc}) can be used for constructing another 
exactly solvable potential. Starting from the first three superpotentials 
(\ref{Wn1osc})
we obtain in the same way as before the next pair of potentials
\begin{eqnarray}\label{Vm2osc}
{\cal V}_-(x)&=&{\epsilon^2 x^2\over 2}+{9\epsilon\over 2},
\\\label{Vp2osc}
{\cal V}_+(x)&=&{\epsilon^2 x^2\over 2}
+{8\epsilon(2\epsilon x^2-3)\over 3+12\epsilon x^2+4\epsilon^2 x^4}
+{384\epsilon^2 x^2\over ( 3+12\epsilon x^2+4\epsilon^2 x^4)^2}
+{7\epsilon\over 2}.
\end{eqnarray}
Repeating this procedure many times we obtain the following pairs of SUSY 
partner potentials
\begin{eqnarray}\label{Vm+oscn}
V_+(n,x)={\cal V}_-(n,x)&=&{\epsilon^2 x^2\over 2}
+\left(2n-{1\over 2}\right)\epsilon.
\\\label{Vmnosc}
V_-(n,x)={\cal V}_+(n,x)&=&{\epsilon^2 x^2\over 2}
+8\epsilon n(2n-1)
{H_{2n-2}(i\sqrt{\epsilon}x)\over H_{2n}(i\sqrt{\epsilon}x)}
\\\nonumber
&-&16\epsilon n^2
\left(
{H_{2n-1}(i\sqrt{\epsilon}x)\over H_{2n}(i\sqrt{\epsilon}x)}
\right)^2+{(4n-1)\epsilon\over 2},
\end{eqnarray}
where $H_n(x)$ --- Hermite polynomial. 
Note that the case $n=1$ in the potential (\ref{Vmnosc}) corresponds to 
(\ref{Vp1osc}) and the case $n=2$ corresponds to (\ref{Vp2osc}).
The potentials $V_-(n,x)$ (\ref{Vmnosc}) are just the special cases of the 
SUSY partner potential of a harmonic oscillator obtained by Sukumar 
\cite{Suk1} and they were previously obtained by Bagrov and Samsonov 
\cite{Bag1,Bag2} via the Darboux method and latter by Junker and Roy 
\cite{Jun} within the SUSY approach.

The application of the same procedure for the Morse and Rosen--Morse 
potentials as well as for radial harmonic oscillator and hydrogen atom 
provides chains of exactly solvable potentials.
All these potentials are just special  
cases of the potentials obtained in \cite{Jun} by Junker and Roy.  
Nevertheless they seem to be interesting
because all of them are expressed in terms of the 
elementary functions only. For example starting from the Morse potential for 
which corresponding superpotentials and distances between the energy 
levels read
\begin{eqnarray}
W_n(x)=\epsilon+1/2-n-e^{-x}, && \epsilon_n=\epsilon-n,
\end{eqnarray}
we construct the function
\begin{eqnarray}
U(x)=4(\epsilon-e^{-x})(\epsilon-1-e^{-x}).
\end{eqnarray}
An explicit calculations show us that we should take function 
${\cal R}_0(x)$ in the form
\begin{eqnarray}\label{cR0Mor}
{\cal R}_0(x)=
{
4e^{-2x}+6(1-2\epsilon)e^{-x}+3(1+4\epsilon(\epsilon-1))
-2\epsilon(1-3\epsilon+2\epsilon^2)e^x
\over
2e^{-x}+(1-2\epsilon)
}
\end{eqnarray}
to reproduce Morse potential. Function ${\cal R}_0(x)$ does not coincide 
with $R(x)$ in this case, the former is negative within the interval 
limited by zeros of the expression (\ref{cR0Mor}).
Following the same procedure as in the case of harmonic oscillator
we obtain such a sequence of potentials
\begin{eqnarray}V_-(0,x)&=&
{(1+2\epsilon)^2\over 8}+{e^{-2 x}-2(\epsilon+1)e^{-x}\over 2},\\
V_-(1,x)&=&
{(1+2\epsilon)^2\over 8}+{e^{-2 x}-2(\epsilon-1)e^{-x}\over 2}\\\nonumber
&+&
{2(\epsilon(2\epsilon-1)e^x-2)\over \epsilon
(2-2(2\epsilon-1)e^x+\epsilon(2\epsilon-1)e^{2x})}
-{8((2 \epsilon-1)e^x-1 )\over 
\epsilon(2-2(2\epsilon-1)e^x+\epsilon(2\epsilon-1)e^{2x})^2},\\
V_-(2,x)&=&
{(1+2\epsilon)^2\over 8}+{e^{-2 x}-2(\epsilon-3)e^{-x}\over 2}
\\\nonumber
&+&
{
(2\epsilon-3)
\left(
8 e^x-
(\epsilon-1)
\left(
48 e^{2x}
-(2\epsilon-1)
\left(
36 e^{3x}
-16\epsilon e^{4x}
\right)   
\right)   
\right)   
\over
4-
(2\epsilon-3)
\left(
8 e^x 
-(\epsilon-1)
\left(
12 e^{2 x}
-(2 \epsilon-1)
\left(
4 e^{3 x}
-\epsilon
e^{4 x}
\right)
\right)   
\right)   
}
\\\nonumber
&+&
{
\left(
(2\epsilon-3)
\left(
8 e^x 
-(\epsilon-1)
\left(
24 e^{2 x}
-(2 \epsilon-1)
\left(
12 e^{3 x}
-\epsilon
4 e^{4 x}
\right)   
\right)   
\right)   
\right)^2
\over
\left(
4-
(2\epsilon-3)
\left(
8 e^x 
-(\epsilon-1)
\left(
12 e^{2 x}
-(2 \epsilon-1)
\left(
4 e^{3 x}
-\epsilon
e^{4 x}
\right)
\right)   
\right)   
\right)^2
}.
\end{eqnarray}
Corresponding SUSY partners are following Morse potentials
\begin{eqnarray}
V_+(n,x)&=&
{(1+2\epsilon)^2\over 8}+{e^{-2 x}-2(\epsilon-2n)e^{-x}\over 2}.
\end{eqnarray}
We can proceed with this procedure as long as necessary.

The most interesting fact is that at each step of the suggested procedure 
for the all mentioned above potentials 
\begin{equation} \label{calHH}
{\cal H}_-=H^{(2)}_++\epsilon+\epsilon_1.
\end{equation}
One can easily check that due to the connection (\ref{11}) and (\ref{Hset})
\begin{eqnarray}
E^{(2)+}_n=E^{(2)-}_{n+1}=E^{(1)+}_{n+1}-\epsilon_1
=E^{+}_{n+2}-\epsilon_1-\epsilon=E^{-}_{n+3}-\epsilon_1-\epsilon.
\end{eqnarray}
and therefore the energy levels of the Hamiltonian ${\cal H}_-$ 
coincide with that of the Hamiltonian $H_-$ saving the three lowest levels 
of the latter which are not present in the spectrum of the former
\begin{eqnarray}
{\cal E}^-_n=E^-_{n+3},&&n=0,1,\ldots
\end{eqnarray}
It gives us immediately the energy spectrum of the new Hamiltonian 
${\cal H}_+$ 
\begin{eqnarray}
{\cal E}^-_0=0,&&
{\cal E}^-_n=E^-_{n+2},\ \ \ \ n=1,2,\ldots
\end{eqnarray}
Besides the connection (\ref{calHH}) allows us to obtain easily 
all the eigenfunctions of the excited states for the new exactly solvable 
potential ${\cal H}_+$. Using relation (\ref{PhiPhi}) we obtain
\begin{eqnarray}\label{Phin} 
\varphi_{n}^+(x)=
C_n^+
\left({d\over dx}+{\cal W}(x)\right)\psi_{n-1}^{(2)+}(x),
&& n=1,2,\ldots 
\end{eqnarray}

Thus for all mentioned exactly solvable potentials we can construct a 
sequence of Hamiltonians $H_n=-{1\over 2}{d^2\over dx^2}+V(n,x)$ 
with the same energy levels as $H_0$ except the $2n$ lowest excited states 
of the latter. All of them possess zero--energy ground state.

\section{Conclusions}
We have obtained a general solution of the set of equations 
(\ref{Wset}) for $N=2$ given by the expressions (\ref{Wp}) and 
(\ref{Wneq}) ($n=0,1$). Thus we can write down explicit expressions
for superpotentials $W_0(x)$, $W_1(x)$ and $W_2(x)$ and afterwards for 
potential $V_-^{(0)}(x)=(W_0^2(x)-W_0'(x))/2$ being just a general 
expression for a QES potential with three explicitly known eigenstates.
General expressions for the corresponding eigenfunctions are also presented 
(\ref{9}), (\ref{Psi1}). The QES potential is expressed in terms of the 
distances $\epsilon$, $\epsilon_1$ between neighbouring energy levels and 
an arbitrary function $U(x)$. To ensure nonsingularity of the potential we 
need to put a number of limitations on the function $U(x)$. Using this 
expression we have obtained some new QES potentials. In special cases our 
potentials reproduce those studied earlier. 

There obviously arises a question 
if the suggested scheme could be generalized to construct QES potentials 
with the number of explicitly known eigenstates large than three.
In such a case we have the set of equations (\ref{Wpset}) consisting of 
$N-2$ equations. Let us remind that $N$ is number of excited states of QES 
potential which we would like to construct.
In order to reduce this set of equations we can proceed with the 
scheme described in section \ref{QESc} (eqs. (\ref{Wset})--(\ref{Wp})).The 
solution of each equation of the set (\ref{Wpset}) can be written down in 
the form (\ref{Wp}), where $W_+(x)$ is replaced by $W_+^{(n)}(x)$,
$\tilde{W}_+(x)$ is replaced by $W_+^{(n+1)}(x)$ 
and $U_n(x)=W_+^{(n)}(x)W_+^{(n+1)}(x)$ is instead of $U(x)$.
Two neighbouring equations of the set (\ref{Wpset}) yield us two 
different expressions for the same function $W_+^{(n)}$ in terms of the 
functions $U_n(x)$ and $U_{n+1}(x)$ correspondingly. Thus we obtain 
following set of equations for functions $U_n(x)$
\begin{eqnarray}  \label{Uset}
{U_n'(x)\left(1+
{\cal R}_n(x)
\right)\over 2 (U_n(x)+2\epsilon_n)}
=
{2 U_{n+1}(x)(U_{n+1}(x)+2\epsilon_{n+1})\over 
U_{n+1}'(x)\left(1+
{\cal R}_{n+1}(x)
\right)},&&n=0,\dots,N-3,
\end{eqnarray}
where
\begin{equation}
{\cal R}_n(x)
=\pm\sqrt{1+4{U_n(x)(U_n(x)+2\epsilon_n)(U_n(x)-2\epsilon_{n+1})\over 
U_n'(x)^2}}.
\end{equation}
Number of equations in this set is one less than number of equation in the 
set (\ref{Wpset}) for functions $W_+^{(n)}(x)$ and correspondingly it is two 
less than the number of the initial equations (\ref{Wset}) for 
surperpotentials. In the case $N=3$ we have just one equation which we need 
to solve to construct QES potentials with four known eigenstates. 
Thus one can see that the obtaining of general expression for QES 
potentials with more than three explicitly known eigenstates is 
essentially more complicated.

Another point is that the suggested scheme allows one to 
construct in a simple way the sequences of SUSY partner potentials of 
exactly solvable ones. At each step we obtain a new exactly solvable 
potential with identical spectrum, apart from missing two the lowest 
excited states. Note that this approach in contrast to 
method used in \cite{Suk,Suk1,Bag1,Bag2,14,Jun} does not 
require knowledge of general solution of corresponding Schr\"odinger 
equation for initial potential. 
It would be interesting to apply the same approach 
to the known QES potentials with explicitly known $N$ eigenstates
to construct new QES potentials with picked out two the lowest excited 
states. But the latter case is more complicated than the case of 
shape--invariant potentials and it will be the subject of a separate paper.


\begin{thebibliography}{18}
\bibitem{Singh} V.~Singh, S.~N.~Biswas, K.~Dutta, Phys. Rev. D {\bf 18} 
            (1978) 1901.
\bibitem{Fless} G.~P.~Flessas, Phys. Lett. A {\bf 72} (1979) 289.
\bibitem{Raz} M.~Razavy, Am. J. Phys. {\bf 48} (1980) 285; Phys. Lett A {\bf 82}
             (1981) 7.
\bibitem{Khare1} A.~Khare, Phys. Lett. A {\bf 83} (1981) 237.

\bibitem{Zas1} O.~B.~Zaslavskii, V.~V.~Ul'yanov, V.~M.~Tsukernik,
             Fiz. Nizk. Temp. {\bf 9} (1983) 511.
\bibitem{Zas2} O.~B.~Zaslavsky, V.~V.~Ul'yanov, Zh. Eksp. Teor. Fiz.
             {\bf 87} (1984) 1724.
\bibitem{Tub1} A.~V.~Turbiner, A.~G.~Ushveridze, Phys. Lett. A {\bf 126} 
             (1987) 181.
\bibitem{Tub2} A.~V.~Turbiner, Commun. Math. Phys. {\bf 118} (1988) 467.
\bibitem{Shif} M.~A.~Shifman, Int. Jour. Mod. Phys. A {\bf 4} (1989) 2897.
\bibitem{Jatkar} D.~P.~Jatkar, C.~Nagaraja Kumar, A.~Khare, Phys. Lett. A 
             {\bf 142} (1989) 200.
\bibitem{Roy1} P.~Roy, Y.~P.~Varshni, Mod. Phys. Lett. A {\bf 6}
             (1991) 1257.
\bibitem{Gang} A.~Gangopadhyaya, A.~Khare, U.~P.~Sukhatme,
             Phys. Lett. A {\bf 208} (1995) 261.
\bibitem{Ush} A.~G.~Ushveridze, {\it Quasi-exactly solvable models in 
	quantum mechanics}, (Institute of Physics Publishing, Bristol 1994).
\bibitem{Bagchi} B.~Bagchi, C.~Quesne, Phys. Lett. A {\bf 230} (1997) 1.
\bibitem{Finkel} F.~Finkel, A.~ Gonz\'{a}lez--L\'{o}pez, 
	M.~A.~Rodr\'{\i}guez, J. Phys. A {\bf 30} (1997) 6879
\bibitem{Ul} V.~V.~Ulyanov, O.~B.~Zaslavskii, J.~V.~Vasilevskaya,
            Fiz. Nizk. Temp. {\bf 23} (1997) 110. 
\bibitem{Vas} Yu.~V.~Vasilevskaya, V.~V.~Ulyanov, Ukr. Fiz. Zh. {\bf 43} 
	    (1998) 363.
\bibitem{Khare2} A.~Khare, B.~P.~Mandal, Phys. Lett. A {\bf 239} (1998) 197.
\bibitem{Bender} C.~M.~Bender, S.~Boettcher, J. Phys. A {\bf 31} 
	(1998) L273.
\bibitem{Suk}  C.~V.~Sukumar, J. Phys. A {\bf 18} 
	(1985), L57.
\bibitem{Suk1}  C.~V.~Sukumar, J. Phys. A {\bf 18} 
	(1985), 2917.
\bibitem{Bag1} V.~G.~Bagrov, B.~F.~Samsonov, Teor. Mat. Fiz. {\bf 104} 
	(1995), 356.
\bibitem{Bag2}  V.~G.~Bagrov, B.~F.~Samsonov, J. Phys. A {\bf 29} 
	(1996), 1011.
\bibitem{Fer1} D.~J.~Fern\'andez~C., M.~L.~Glasser, L.~M.~Nieto,
	Phys. Lett. A {\bf 240} (1998) 15.
\bibitem{Fer2} D.~J.~Fern\'andez~C., V.~Hussin, B.~Mielnik,
	Phys. Lett. A {\bf 244} (1998) 309.
\bibitem{Lahiri} A.~Lahiri, P.~K.~Roy and B.~Bagchi, Int. J. Mod. Phys. A
{\bf 5} (1990) 1383.
\bibitem{13} F.~Cooper, A.~Khare, U.~Sukhatme, Phys. Rep. {\bf 251} 
(1995) 267.
\bibitem{QES2S} V.~M.~Tkachuk, Phys. Lett. A {\bf 245} (1998) 177 [preprint 
		quant-ph/9801021 (1998)].
\bibitem{SUSY} L.~E.~Gendenshteyn, Pisma Zh. Eksp. Teor.Fiz. {\bf 
	38} (1983) 299.
\bibitem{15} A. de Souza Dutra, Phys. Rev. A {\bf 47} (1993) R2435.
\bibitem{Nik} J.~Beckers, N.~Debergh, A.~G.~Nikitin, Mod. Phys. Lett. A 
{\bf 8} (1993) 435.

\bibitem{14} P.~Roy, G.~Junker, Phys. Lett. A {\bf 232} (1997) 155.
\bibitem{Jun} G. Junker, P. Roy, preprint quant-ph/9803024 (1998).
\end{thebibliography}
\end{document}